# On-Chip Spin-Orbit Controlled Excitation of Quantum Emitters Coupled to Hybrid Plasmonic Nanocircuits

Y. H. Kan[1], Shailesh Kumar[2], Fei Ding[2], C. Y. Zhao[1]*, Sergey I. Bozhevolnyi[2]*

[1]Institute of Engineering Thermophysics, Shanghai Jiao Tong University, Shanghai, 200240, China

[2]Centre for Nano Optics, University of Southern Denmark, DK-5230 Odense M, Denmark

**Abstract.** On-chip realization of complex photonic functionalities is essential for further progress in planar integrated nanophotonics, especially when involving nonclassical light sources such as quantum emitters (QEs). Hybrid plasmonic nanocircuits integrated with QEs have been attracting considerable attention due to the prospects of significantly enhancing QE emission rates and miniaturizing quantum nanophotonic components. Spin-orbit interactions on subwavelength scales have been increasingly explored in both conventional and quantum nanophotonics for realization and utilization of the spin-dependent flow of light. Here, we propose and realize a dielectric-loaded plasmonic nanocircuit consisting of an achiral spin-orbit coupler for unidirectional routing of pump radiation into branched QE-integrated waveguides. We demonstrate experimentally the circular-polarization controlled coupling of 532-nm pump laser light into polymer-loaded branched waveguides followed by the excitation of spatially separated (by a distance of ~ 10 $\mu$m) QEs, nanodiamonds, with multiple nitrogen vacancy centres, that are embedded in and efficiently coupled to the corresponding waveguides. The realization of on-chip spin-orbit controlled excitation of different QEs coupled to branched waveguides opens new avenues for designing complex quantum plasmonic nanocircuits exploiting the spin degree of freedom within chiral quantum nanophotonics.

**Introduction**

Integrated photonic and hybrid plasmonic configurations consisting of waveguide nanocircuits coupled with quantum emitters (QEs) have attracted considerable attention due to potential applications in emerging quantum information technologies, such as communication, computation, imaging and sensing[1-8]. Different kinds of plasmonic waveguides, including metal wedges, V-grooves, single and parallel nanowires, have been considered for QE coupling to strongly confined plasmon modes in the form of propagating surface plasmon polaritons (SPPs)[9-18]. Even though these waveguides are able of supporting extremely strongly confined SPP modes, their suitability for deterministic, accurate and practical integration with QEs is rather challenging, if possible at all, due to their complicated fabrication involving chemical synthesis (for metallic nanowires[12-14]), focused ion-beam milling (for V-shaped channels[15,16]) and other sophisticated fabrication/assembly techniques[17-19]. Dielectric loaded surface plasmon polariton waveguides (DLSPPWs) can be considered as a promising integration platform to overcome this issue, because the top–down lithography-based fabrication techniques make it easier to embed preselected QEs inside the dielectric waveguides[20-23]. Besides, DLSPPWs exhibit also a relatively large propagation length without suffering from larger losses in short wavelength[24-28]. Considering the figure of merit (FOM), which is defined by the product of QE decay rate enhancement (Purcell factor), SPP mode propagation length, and the QE–SPP mode coupling efficiency (β-factor), DLSPPWs were found to show superior performance compared to V-grooves, wedge waveguides, and nanowires[20,21]. Remarkably, DLSPPWs have also been found suitable for the on-chip remote QE excitation[29], opening thereby an exciting perspective for designing on-chip integration of solid-state photonic systems[30].

Spin-orbit interactions (SOI) of light on the subwavelength scales that couple the polarization and spatial degrees of freedom have recently attracted a great deal of attention due to bringing in novel functionalities to optical nano-devices[31-36]. The spin-selective flow of light has been demonstrated with various photonic[37-39] and plasmonic[40-42] configurations. Very recently, the SOI have also been exploited to realize the on-chip electrical detection of the spin state of incident photons[43]. Combining the SOI control with on-chip remote QE excitation seems attractive and promising from the viewpoint of developing complex quantum nanocircuits exploiting the spin degree of freedom, even though its practical realization poses several formidable challenges by requiring the identification of waveguide circuit configuration that would be suitable for SOI-controlled excitation and also amenable to deterministic and accurate integration with individual QEs.

In this work, we propose a dielectric-loaded plasmonic nanocircuit, consisting of an achiral spin-orbit coupler (SOC) and branched QE-integrated waveguides, to realize SOI-controlled excitation of remote QEs in a scalable on-chip implementation. First, using full-wave simulations we design the achiral SOC for selectively and unidirectionally routing the SPP flow into two DLSPPW branches with a large contrast (~ 30 times) for right and left circular polarization (CP) incidence. We then fabricate SOC-DLSPPW nanocircuits with different DLSPPW lengths to characterize the SOC performance as well as to ascertain the DLSPPW mode propagation length and verify efficient long-range energy transfer. By varying the state of polarization of the incident beam, we demonstrate that the fabricated SOC-DLSPPW nanocircuits exhibit the expected performance in both CP-controlled unidirectional coupling and DLSPPW mode propagation. To realize experimentally the CP-controlled excitation of QEs, two spatially separated nanodiamonds (NDs) containing multiple nitrogen vacancy (NV) centres are first identified, and their coordinates in the reference frame are determined. The previously worked out SOC-DLSPPW design is then used for precisely and symmetrically aligning and fabricating the hydrogen silsesquioxane (HSQ) based SOC-DLSPPW nanocircuit. Using subsequent optical characterization, we demonstrate, for the first time to our knowledge, that the spatially separated (by a distance of ~ 10 $\mu$m) QEs can be selectively and remotely excited using an incident pump beam with different circular polarizations. The realization of on-chip SOI-controlled excitation of different QEs coupled to branched waveguides demonstrates great application potential of this novel hybrid plasmonic configuration and opens new avenues for designing complex quantum plasmonic nanocircuits exploiting the spin degree of freedom within chiral quantum nanophotonics.

**Results**
**Hybrid nanocircuit for spin-orbit control of QE excitation.**
The proposed hybrid nanocircuit for on-chip spin-orbit controlled QE excitation consists of the achiral SOC for unidirectional routing of pump radiation into one of the two branched QE-integrated DLSPPWs (depending on the spin of pump photons), with the corresponding QE excited and emitting into the DLSPPW (Fig. 1a). Upon excitation with a 532-nm left CP beam, the achiral SOC would, in the ideal case of perfect design, couple the light into the DLSPPW mode propagating in the right branch. The DLSPPW mode propagating along the branch would then excite the QE embedded in the DLSPPW. Following the QE excitation, its emission couples to the DLSPPW mode[20] and propagates along the DLSPPW, reaching its termination with the designed out-coupling grating that couples the mode out into a free propagating radiation (Fig. 1a). The hybrid plasmonic SOC-DLSPPW nanocircuit represents an HSQ ridge structure fabricated on a silver-coated silicon substrate (Fig. 1b). The SOC configuration comprises a 250-nm-diameter nanodisk and a concentric half-ring with the inner and outer radii of 250 nm and 500 nm, respectively (inset in Fig. 1b). The 180-nm-high and 250-nm-wide DLSPPWs are branched out of the SOC. Two remote (at a distance of ~ 10 $\mu$m) NDs with multiple NV centres, representing the QEs, are integrated each in the corresponding DLSPPW at a distance of ~ 2 $\mu$m away from the out-coupling 550-nm-period gratings. The fabricated nanocircuits are optically characterized in an experimental setup allowing one to image and monitor both the scattered incident pump radiation and QE emission (see Methods for details).

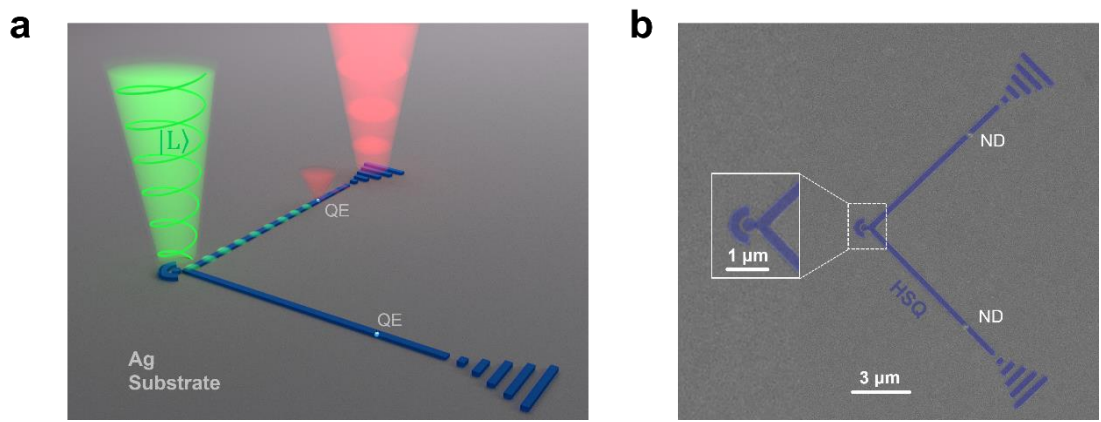

**Fig. 1 On-chip spin-orbit controlled QE excitation in plasmonic nanocircuits. a**

Schematic depicting a 532-nm left CP beam unidirectionally coupled into the specific DSPPW mode by SOI in an achiral SOC, selectively exciting the corresponding QE embedded in the DLSPPW. The QE emission couples to the DLSPPW mode propagating towards the DLSPPW termination with the designed out-coupling grating. **b** Colorized scanning electron microscopy (SEM) image showing the top view of the nanocircuit. The inset represents the magnified SEM image of the achiral SOC.

**Modelling of the SOC.**

We first conduct simulations using finite-difference time-domain (FDTD) method to model the SOI controlled coupling by SOI in the proposed SOC, which is illuminated with a normally incident Gaussian beam at the pump wavelength $\lambda_0$ = 532 nm. It is observed that, depending on whether left CP (LCP) or right CP (RCP) illumination is used, one or another DLSPPW mode is excited at the SOC and subsequently propagates unidirectionally in the corresponding DLSPPW (Fig. 2a). The insets present the cross-section field intensity distributions in two branching DLSPPWs, showing well-confined DLSPPW modes with a large contrast in the excitation efficiency of these DLSPPWs. The DLSPPW mode intensities as a function of the incident light polarization controlled by the polarization angle (see Methods for details) are seen to vary accordingly: from equal intensities for the linear polarization (LP) incidence (the polarization angle of 0, $\pi$, and $2\pi$) to strongly contrasting intensities for the RCP (the polarization angle of $\pi/2$) and LCP (the polarization angle of $3\pi/2$) illumination (Fig. 2b). For the CP incidence, the SOI controlled coupling favours the DLSPPW mode propagation in one particular DLSPPW, while only a small fraction of radiation is directed into the other one, resulting in a directionality contrast of ∼ 30 (Fig. 2c).

The observed spin-selective energy flow can be considered as manifestation of the coupling of the spin and orbital angular momentum of light, i.e., the SOI, on subwavelength scales[38,39]. The direction of the transverse spin angular momentum associated with evanescent fields is uniquely locked with the propagation direction of SPP modes, implying that the SPP modes with opposite wavevectors carry opposite transverse spins[35,42]. The spin of the incident photon interacts with the inherent transverse spin of the SPP mode propagating along the outer edge of the nanodisk that functions as a subwavelength scatterer enabling momentum matching for coupling to the SPP modes. Consequently, the propagation direction of the edge SPP mode (launching the corresponding DLSPPW mode) is determined by the helicity of the incident light matching the handedness of the SPP mode evanescent tail (see Supplementary Note 1 for configuration design and optimization details). The SOI nature of spin-controlled coupling in the considered configuration is intrinsically broadband (Fig. 2c) and thereby very robust with respect to fabrication imperfections.

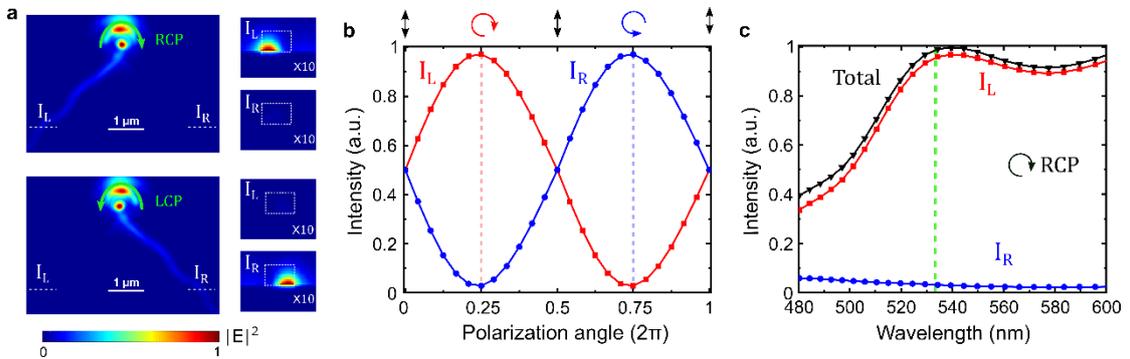

**Fig. 2 Circular-polarization controlled coupling by SOI in the designed nanoscale directional coupler. a** Simulated near-field intensity distribution for the illumination

with RCP (top panel) and LCP (bottom panel) beams, demonstrating the unidirectional radiation routing into the left and right DLSPPWs, respectively. The dashed white lines are the positions of monitors used to monitor the transmission of DLSPPW modes through the waveguides. The insets are the field intensity distributions of the DLSPPW cross-section profiles. **b** Output intensities of the two DLSPPWs as a function of the polarization angle with the angles of 0, $\pi$, and $2\pi$ corresponding to LP, while those of $\pi/2$ and $3\pi/2$ to the RCP and LCP incidence, respectively. **c** The DLSPPW mode intensities as a function of the wavelength for the RCP beam incidence. The green dashed line marks the pump laser wavelength of 532 nm.

**Experimental characterization of the SOC-DLSPPW nanocircuits.**

To experimentally characterize the SOC performance and ascertain the DLSPPW mode propagation length in SOC-DLSPPW nanocircuits at the wavelength of 532 nm, we conduct the comprehensive optical characterization of nanocircuits fabricated (without QEs being integrated) on a 250-nm-thick silver film deposited on a silicon wafer (Fig. 3a), featuring DLSPPWs of different lengths ($L$ = 2, 4, 6, 8, 10, and 12 $\mu$m). By adjusting the orientation of a quarter-wave plate with respect to that of a linear polarizer (see Methods for details), we produce RCP and LCP as well as LP incident beams (see Supplementary Note 3 for details of the experimental setup). We then characterize the performance of the fabricated SOC-DLSPPW nanocircuit with the DLSPPW length $L$ = 6 $\mu$m by monitoring the radiation scattered by two out-coupling gratings for the RCP, LCP, and LP incidence (Fig. 3b), verifying the occurrence of stark contrast in the out-coupled radiation for the RCP and LCP cases. At the same time, for the LP incidence, both DLSPPWs branched out of the SOC are apparently equally excited resulting in roughly equal amount of scattering from the out-coupling gratings (Fig. 3b).

To gain a deeper insight into SOI mechanisms at work in the fabricated SOC and compare the SOC performance with the simulation results (Fig. 2), we measure the out-coupled light intensities for the two DLSPPWs with the state of polarization of the incident beam being gradually changed (between the LP, RCP, LCP, and intermediate elliptical polarizations) by rotating the quarter-wave plate from 0° to 360° (Fig. 3c). The out-coupled intensities exhibit the $\pi$-shifted harmonic behaviour expected for the SOI-controlled coupling with the SOC-DLSPPW combination playing the role of an analyser for CP light, confirming that the directionality of DLSPPW mode propagation in our device is exclusively determined by the SOI. On average, the contrast of intensities at DLSPPW out-coupling gratings measured for the RCP and LCP incidence amounts to 32, which is in good agreement with the simulation results.

The DLSPPW mode propagation length is determined by comparing the attenuation of scattered light intensities at the out-coupling gratings for different DLSPPW lengths. The data obtained by treating the optical images of fabricated SOC-DLSPPW nanocircuits of different length (Fig. 3a) are found to be fitting well to an exponential length-dependent decay of the out-coupled light intensities, resulting in the DLSPPW mode propagation length of $6.8 \pm 2.4\,\mu$m (Fig. 3d). This propagation length is understandingly considerably shorter than in the previously reported on-chip QE excitation realized with DLSPPWs on monocrystalline silver platelets[30], but is definitely large enough for realizing remote QE excitation. The knowledge of the DLSPPW mode propagation length is very important for the selection of appropriately separated NDs in view of their integration in the SOC-DLSPPW nanocircuits.

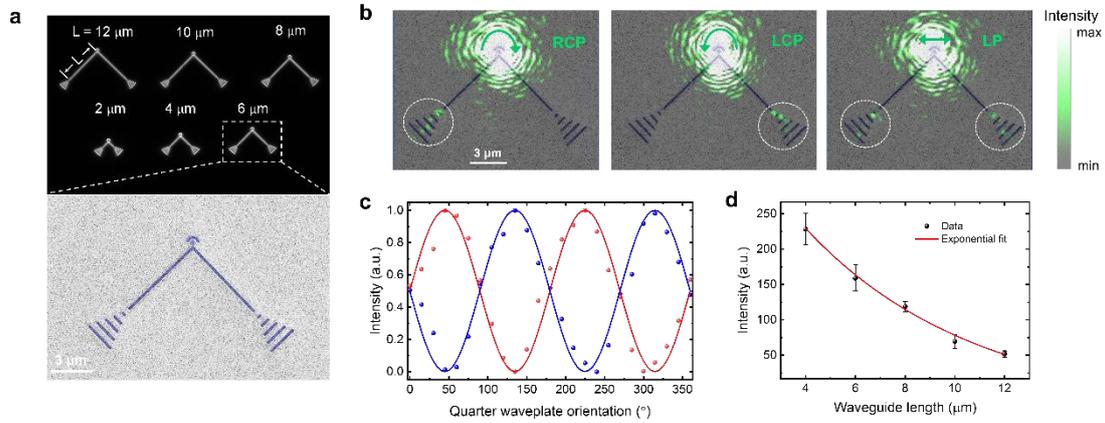

**Fig. 3 Experimental demonstration of spin-sorting in SOC-DLSPPW nanocircuits. a** Dark-filed microscopy image of the fabricated nanocircuits with different DLSPPW lengths *L* (top panel) and the SEM image (bottom panel) of the nanocircuit based on the DLSPPWs with the length *L* = 6 $\mu$m. **b** Visualization of CP-controlled excitation of the SOC-DLSPPW nanocircuit (*L* = 6 $\mu$m) with the RCP (left panel), LCP (middle panel), and LP (right panel) light incidence at the wavelength of 532 nm. The optical images are superimposed with the device configuration. **c** Light intensities evaluated at two out-coupling gratings of the SOC-DLSPPW nanocircuit (*L* = 6 $\mu$m) as a function of the quarter-wave plate orientation with respect to the polarizer. The state of polarization of the excitation laser beam changes continuously between LP (0, 90°, and 180°), RCP and LCP (45° and 135°, respectively). **d** Attenuation of scattered light intensities at the out-coupling gratings for different DLSPPW lengths (black squares) and exponential fitting curve (red line), resulting in the DLSPPW mode propagation length of $6.8 \pm 2.4$ $\mu$m at the wavelength of 532 nm.

**On-chip CP-controlled QE excitation.**

To unambiguously demonstrate the capability of the proposed hybrid plasmonic nanocircuit of on-chip CP-controlled excitation of remote QEs, we experimentally assemble a SOC-DLSPPW nanocircuit, in which two preselected distant (~ 10 $\mu$m) NDs are individually integrated inside each DLSPPW. Here, we use NDs with multiple NV centres as the QEs, being motivated by their remarkable optical characteristics including brightness and room temperature stability[44-46]. The positions of selected NDs are determined using dark-field images of NDs spin-coated on a silver film with prefabricated alignment markers, which enable precise and symmetrical ND integration in the corresponding DLSPPW branches at a distance of ~ 2 $\mu$m away from the out-coupling 550-nm-period gratings (see Methods and Supplementary Note 2 for details). This distance is sufficiently large to distinguish the NV emission coming directly from the NDs (caused by the NV coupling to free propagating optical fields) and that coming from the out-coupling gratings, being caused by the NV coupling to the DLSPPW mode propagating toward the gratings.

With the RCP or LCP (532-nm-wavelength) pump laser beams incident on the SOC, the corresponding DLSPPW modes are selectively excited, propagating in each branch toward the respective NDs. We perform a wide-field collection fluorescence imaging around the grating areas by galvanometric mirror scanning. The fluorescence emission is detected at the focal plane, using a Fourier lens before a charged coupled device (CCD) camera (see Methods and Supplementary Note 3 for details). Bright spots from the embedded NDs due to NV emission are seen to be located inside the DLSPPWs, confirming the remote CP-controlled NV excitation (Fig. 4a). Remarkably, one can also observe the NV emission that is scattered by the out-coupling gratings, demonstrating the fact of the NV emission coupling into propagating DLSPPW modes. It should be borne in mind that due to DLSPPW mode propagation losses and limited collection efficiency of the scattered (by the out-coupling gratings) light, the NV emission originated at the out-coupling gratings is expected to be weaker than the direct emission[30]. Note that the emitters used in this experiment are of different brightness, a circumstance that results in visually different contrasts between fluorescence image of the two DLSPPW branches for different CP light incidences. The main reason for this imperfection is due to the different sizes of the selected NDs that also contain different number of NV centres and of different orientations, making the right emitter brighter than the left one under the same intensity of incident light.

In order to quantify the contrast between the emission spots for different pump beam polarizations, line-cuts (along two NDs) of the fluorescence maps are extracted showing the emission intensities as a function of the distance along the cut crossing the DLSPPWs, denoted as the *x*-axis (Fig. 4b). The intensity profiles across the two bright spots under the RCP (blue curve) and LCP (red curve) illuminations demonstrates their strong contrast, manifesting thereby the experimental realization of the on-chip CP-controlled excitation of remote QEs. These emission intensity profiles also show somewhat stronger undesirable signals from the right emitter as noted above (and for the same reasons). Finally, we confirm that the emission from the selected NDs corresponds to that expected from the NV centres by measuring their fluorescence spectra directly and verifying the occurrence of characteristic broadband emission with two peaks corresponding to NV$^-$ and NV$^0$ centres (Fig. 4c), which is the fingerprint of the NV emission from NDs[44-46].

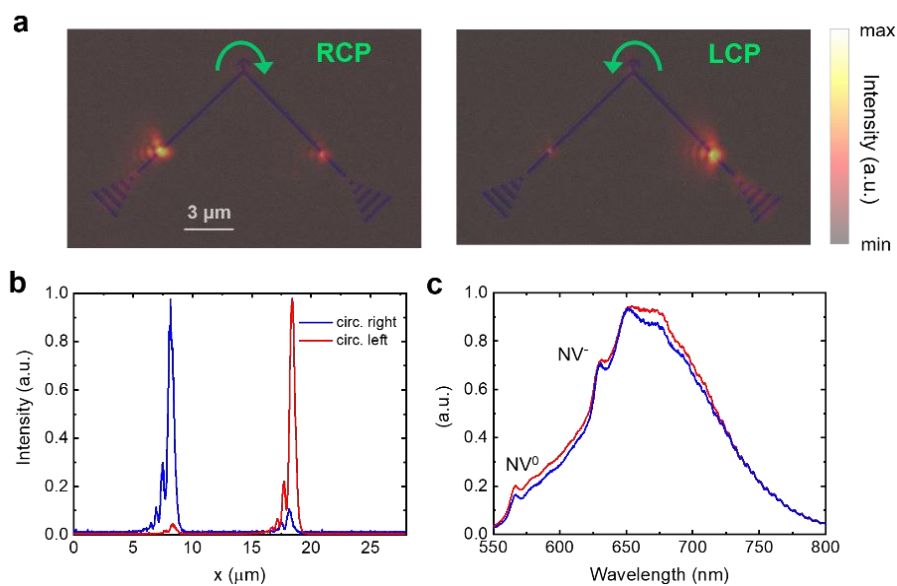

**Fig. 4 On-chip CP-controlled excitation of remote NDs. a** Fluorescence image of the fabricated SOC-DLSPPW nanocircuit showing emission from the NDs containing multiple NV centres remotely and selectively driven with the RCP (right panel) and LCP (left panel) pump laser beams. The emission images are superimposed with the device configuration. **b** Intensity profiles across two bright emission spots in the fluorescence images obtained for the RCP (blue curve) and LCP (red curve) incidence. **c** Fluorescence spectra measured with the NDs used, featuring NV$^-$ and NV$^0$ corresponding peaks typical for the NV emission from NDs.

**Discussion**

On-chip realization of complex photonic functionalities is essential for further progress in planar integrated quantum nanophotonics with SOI being progressively wider explored for realization and utilization of spin-dependent flow of light.[2] In the field of hybrid plasmonic quantum systems, the remote on-chip excitation of QEs as well as the coupling of the QE emission into confined DLSPPW modes have recently been demonstrated[20,21,30]. Our work bridges the gap between DLSPPW-based quantum circuits and chiral quantum nanophotonics by developing and experimentally demonstrating the DLSPPW-based configuration for on-chip SOI controlled excitation of remote QEs. Two major challenges had to be tackled for succeeding in this work: design of SOC configuration capable of operating at visible wavelengths and precise deterministic integration of SOC-DLSPPW circuitry and preselected distant QEs. The first challenge associated with the dissipative nature of plasmonic waveguides has been circumvented by designing the DLSPPW-based SOC configuration, which is to design an achiral SOC that can work at visible wavelengths based on SOI. The second challenge has been dealt with by complementing the previously developed approach for accurately fabricating DLSPPW elements to embed a preselected ND[20] with the possibility for the coordinate transformation (translation and rotation) when matching the SOC-DLSPPW design with the actual positions of two selected NDs.

The realization of on-chip SOI-controlled excitation of different QEs coupled to branched waveguides demonstrates great application potential of this novel hybrid plasmonic configuration and opens new avenues for designing complex quantum plasmonic nanocircuits exploiting spin degree of free don within chiral quantum nanophotonics. Among the next immediate steps to be taken is the usage of various QEs representing single-photon sources, such as quantum dots[11] and NDs with single NV[20,21,47], germanium vacancy (GeV)[30,48] or silicon vacancy (SiV)[49] centres.

Furthermore, one could also introduce nanoantennas or nanocavities near QEs[45] for considerably enhancing (speeding up) the QE emission by the Purcell effect. By integrating different Purcell-enhanced QEs into different waveguide branches, one would develop a versatile quantum nanophotonic platform for realizing spin-selective control of single-photon emission at different wavelengths that represents a rather complex photonic functionality with many potential applications in quantum information technologies.

## Methods
### Sample fabrication.

Two different kinds of samples are fabricated in this work. Both of them start from the ohmic evaporation of a silver film of 250 nm thickness on a Si wafer as substrate, on which gold markers are then made for alignment (See details in the Supplementary Information). For the first fabrication, i.e. dielectric nanocircuits without the integration with NDs, an HSQ layer is then spin-coated at 1200 rpm (60 s) on the sample and prebaked at 160°C for 2 min to make a 180 nm film on the silver surface. The HSQ nanocircuits with different lengths are patterned by electron beam lithography, followed by development using 25% TMAH (Tetramethylammonium hydroxide) for 4 min. For the second fabrication, i.e. hybrid nanocircuits integrated with NDs, 100 nm NDs containing ~400 NV-centres (Adamas technology) are first spin-coated on the samples before HSQ layer spin-coating. The relative positions of the NDs in the coordinate frame of the alignment marks are determined by a dark-field microscope image. The two NDs with a distance around 10 μm are selected and their positions are obtained by fitting two Gaussians to the diffraction-limited spot of the ND (See details in the Supplementary Information). An HSQ layer is subsequently spin-coated and prebaked as the first experiment. The nanocircuits are patterned according to the position of preselected NDs.

### Device characterization.

The pump laser is a linearly polarized 532-nm continuous wave (Crystal laser). For turning the polarization, a broadband quarter-wave plate mounted on a motorized rotation stage is flipped into the optical path together with a polarizer. The beam is focused onto the sample using a ×100 NA 0.9 objective. For the nanocircuit without NDs, the optical fields scattered by the out-coupling gratings are observed and measured by recording the images using a CCD camera. When rotating the quarter-wave plate with respect to the polarizer and thereby changing the angle between their axes from 0° to 360°, the spin-dependent out-coupled light powers are measured. The propagation characteristics of the nanocircuits for several samples of different lengths are measured by comparing the attenuation of the signal. Fluorescence collected by same objective is filtered from the laser light. Recording the fluorescence photon rate with an avalanche photodiode (APD) while scanning the sample, using a piezo stage, allowed for locating two NDs by the recording of fluorescence maps. We have performed wide-field collection fluorescence imaging around our confocal excitation spot and projected the Fourier plane onto a CCD camera. The detailed information about the setup and measurement can be found in the Supplementary Information.

**Numerical Simulations.**
Numerical simulations in this work are performed by commercial FDTD software package (FDTD Solutions, Lumerical Solutions). The optical constants of the silver are taken from Palik's handbook[50]. The refractive index of HSQ is set as 1.41. In order to simulate coupling of differently polarized incident light, two coherent and identically shaped 2-$\mu$m-diameter Gaussian source beams are employed. These optical beams are linearly polarized with variable angle between their polarizations, while their phase difference is maintained constant with the phase retardance of 90°. The resulting incident beam is thereby linearly polarized for the polarization angle of 0° and 180°, while becoming circularly polarized for the polarization angles of 90° and 270° that result in LCP and RCP beams, respectively. To identify the coupled light into each waveguide, we set two frequency-domain profile and power monitors across two waveguides at 3 μm away from the coupler.

## Corresponding Author

*(C.Y.Z.) E-mail: changying.zhao@sjtu.edu.cn

*(S.I.B.) E-mail: seib@mci.sdu.dk

## Acknowledgements

C.Y.Z. and Y.H.K. acknowledge the support from the National Natural Science Foundation of China (Grants No. 51636004) and the China Scholarship Council (No.201806230179). The authors gratefully acknowledge financial support from the European Research Council, Grant 341054 (PLAQNAP). S.I.B. acknowledges the support from the Villum Kann Rasmussen Foundation (Award in Technical and Natural Sciences 2019). F.D. acknowledges the supporting of Villum Experiment (Grant No. 00022988) from Villum Fonden.

## Author contributions

Y.H.K. and S.I.B. conceived the configuration geometry. Y.H.K performed theoretical modelling and the device fabrication. Y.H.K. and S.K. carried out the experimental characterization. Y.H.K., S.K., F.D. and S.I.B. analysed the data. Y.H.K. wrote the manuscript with contributions from all other authors. S.I.B. and C.Y.Z. supervised the project.

## Competing interests

The authors declare no competing interests.

## Additional information

Supplementary information is available for this paper.